\documentclass[aps,prx,10pt,twocolumn,amsmath,showpacs,superscriptaddress,reprint]{revtex4-2} 
\usepackage[hmargin=2cm,vmargin=2.5cm]{geometry}

\usepackage{amsmath}
\usepackage{amsfonts}
\usepackage{bm}
\usepackage{verbatim}
\pdfoutput=1

\usepackage[colorlinks,plainpages=false,linkcolor=blue,urlcolor=blue,citecolor=blue,pdfpagemode=UseNone,pdfstartview=FitBH]{hyperref} 
\usepackage[Charter,greekuppercase=italicized]{mathdesign} 
\usepackage{graphicx}    
\graphicspath{{./}{Figs/}}
\DeclareGraphicsExtensions{.eps,.png,.pdf,.jpg}

\newcommand{\cuox}{[(C$_{2}$H$_{5}$)$_{3}$NH]$_{2}$Cu$_{2}$(C$_{2}$O$_{4}$)$_{3}$}

\begin{document}

\title{Gapless spinons and a field-induced soliton gap in the hyper-honeycomb Cu oxalate framework compound \cuox}							 
\author{C.~Dissanayake}
\affiliation{Department of Physics, University of Central Florida, Orlando, Florida 32816, USA}
\author{A.~C.~Jacko}
\email[Deceased 17th March 2022]{}
\affiliation{School of Mathematics and Physics, The University of Queensland, Brisbane, Queensland 4072, Australia}	
\author{K.~Kumarasinghe}
\author{R.~Munir}
\author{H.~Siddiquee}
\affiliation{Department of Physics, University of Central Florida, Orlando, Florida 32816, USA}
\author{W.~J.~Newsome}
\affiliation{Department of Chemistry, University of Central Florida, Orlando, Florida 32816, USA}
\author{F.~J.~Uribe-Romo}
\affiliation{Department of Chemistry, University of Central Florida, Orlando, Florida 32816, USA}
\affiliation{REACT: Renewable Energy and Chemical Transformations Cluster, University of Central Florida, Orlando, Florida 32816, USA}
\author{E.~S.~Choi}
\affiliation{National High Magnetic Field Laboratory, Florida State University, Tallahassee, Florida 32310, USA}
\author{S.~Yadav}
\altaffiliation[Present address: ]{Walmart Global Tech, Sunnyvale, California 94086, USA}
\affiliation{Department of Physics, University of Florida, Gainesville, Florida 32611, USA}
\author{X.-Z.~Hu}
\altaffiliation[Present address: ]{Institute of Physics, Chinese Academy of Sciences, Beijing 100190, China}
\affiliation{Department of Physics, University of Florida, Gainesville, Florida 32611, USA}
\author{Y.~Takano}
\affiliation{Department of Physics, University of Florida, Gainesville, Florida 32611, USA}
\author{S.~Pakhira}
\affiliation{Ames National Laboratory, Iowa State University, Ames, Iowa 50011, USA}
\affiliation{Institute for Quantum Materials and Technologies, Karlsruhe Institute of Technology, D-76021 Karlsruhe, Germany}
\author{D.~C.~Johnston}
\author{Q.-P.~Ding}
\author{Y.~Furukawa}
\affiliation{Ames National Laboratory, Iowa State University, Ames, Iowa 50011, USA}
\affiliation{Department of Physics and Astronomy, Iowa State University, Ames, Iowa 50011, USA}
\author{B.~J.~Powell}
\affiliation{School of Mathematics and Physics, The University of Queensland, Brisbane, Queensland 4072, Australia}	

\author{Y.~Nakajima}
\email[Corresponding author: ]{Yasuyuki.Nakajima@ucf.edu}
\affiliation{Department of Physics, University of Central Florida, Orlando, Florida 32816, USA}


\date{\today}

\begin{abstract}
  We report a detailed study of the specific heat and magnetic susceptibility of single crystals of a spin liquid candidate: the hyper-honeycomb Cu oxalate framework compound \cuox. The specific heat  shows no anomaly associated with a magnetic transition at low temperatures down to $T\sim$ 180 mK in zero magnetic field. We observe a large linear-in-$T$ contribution to the specific heat $\gamma T$, $\gamma = 98(1)$ mK/mol K$^{2}$, at low temperatures, indicative of the presence of fermionic excitations despite the Mott insulating state. The low-$T$ specific heat is strongly suppressed by applied magnetic fields $H$, which induce an energy gap, $\Delta(H)$, in the spin-excitation spectrum. We use the four-component relativistic density-functional theory (DFT) to calculate the magnetic interactions, including the Dzyaloshinskii-Moriya antisymmetric exchange, which causes an effective staggered field acting on one copper sublattice. The magnitude and field dependence of the field-induced gap, $\Delta(H) \propto H^{2/3}$, are accurately predicted by the soliton mass calculated from the sine-Gordon model of weakly coupled antiferromagnetic Heisenberg chains with all parameters determined by our DFT calculations. Thus our experiment and calculations are entirely consistent with a model of \cuox\ in which anisotropic magnetic exchange interactions due to Jahn-Teller distortion cause one copper sublattice to dimerize, leaving a second sublattice of weakly coupled antiferromagnetic chains. We also show that this model quantitatively accounts for the measured temperature-dependent magnetic susceptibility. Thus \cuox\ is a canonical example of a one-dimensional spin-1/2 Heisenberg antiferromagnet and not a resonating-valence-bond quantum spin liquid, as previously proposed.
\end{abstract}


\maketitle

\section{Introduction}
Frustrated quantum magnets have attracted great interest because of their rich ground states \cite{balen10a,savar17}. For instance, one of the most interesting ground states is the Kitaev spin liquid, which harbors itinerant Majorana fermions and localized visons \cite{kitae06}. This intriguing spin-liquid state can be stabilized in tricoordinate lattices with bond-dependent Ising interactions, including honeycomb and hyper-honeycomb lattices with strong spin-orbit coupling. Extensive theoretical and experimental studies have been conducted to discover materials exhibiting quantum spin-liquid states \cite{zhou17,broho20}.

Recently, metal-organic frameworks (MOFs) have been pointed out to be an excellent platform for such exotic spin-liquid states. MOFs are crystalline materials consisting of inorganic building units containing metal ions and organic linkers \cite{furuk13}. The versatile choice of the building units and linkers potentially provides MOFs with a significant advantage over inorganic materials in stabilizing quantum spin liquids, including the Kitaev spin liquid. Indeed, recent theoretical work predicts that the Ru oxalate Ru$_{2}$(C$_{2}$O$_{4}$)$_{3}$ \cite{yamad17,yamad17a} with the 8$^{2}$.10a-structure \cite{wells}, or the {\bf lig} net in O’Keeffe’s three-letter reticular chemistry structure resource (RCSR) codes \cite{okeef08}, becomes a Kitaev spin liquid. 

\begin{figure}[t]
\includegraphics[width=0.95\linewidth]{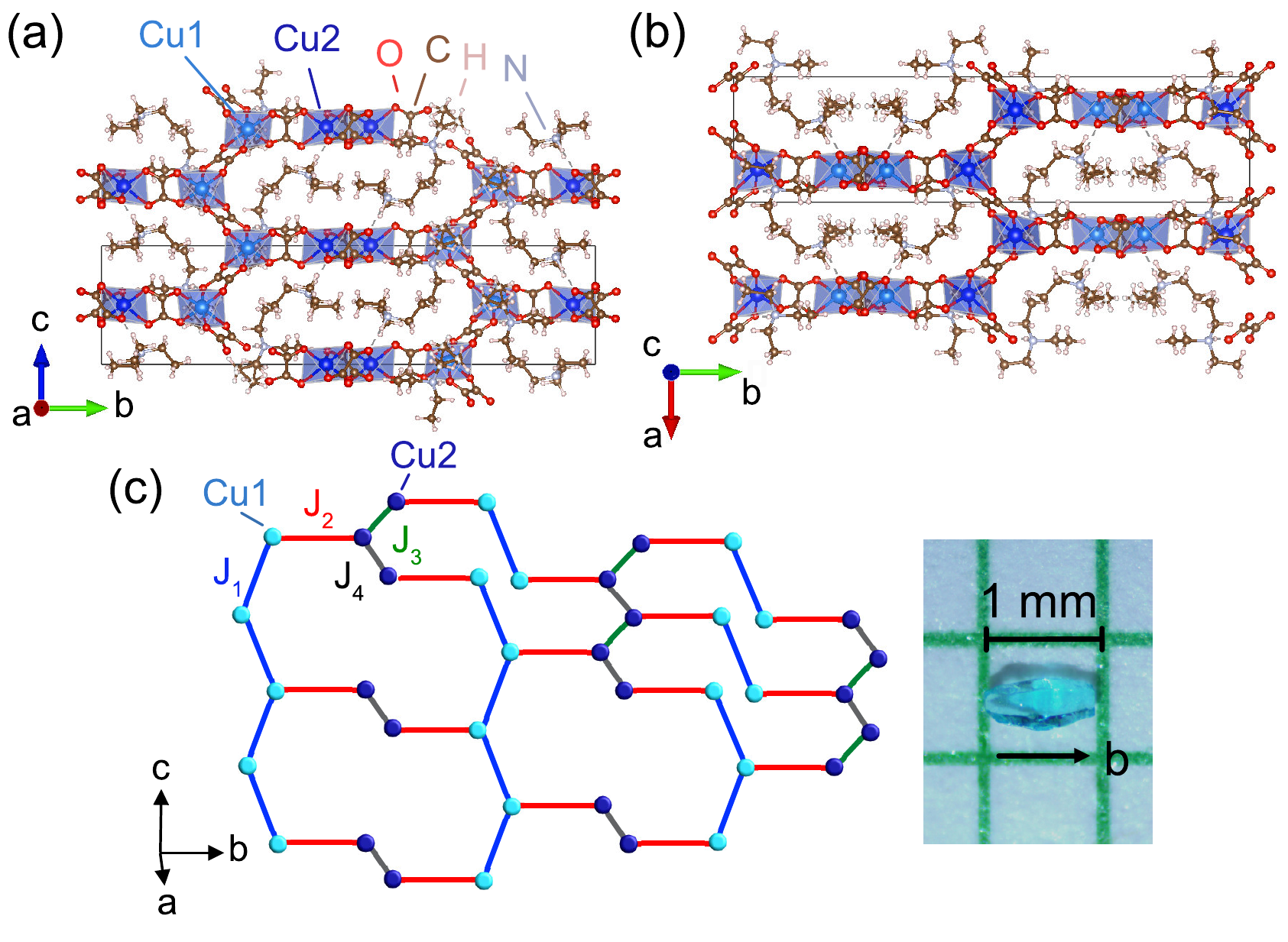}
\caption{\label{fig:network} Crystal structure of \cuox\ \cite{momma11}. Views are shown along (a) the $a$ axis and (b) the $c$ axis. Cu1 ions form chains along the $c$ axis, while Cu2 ions form chains along the $a$ axis. (c) Schematic diagram of Cu network in \cuox\, forming a hyper-honeycomb structure, also known as the {\bf ths} net in O’Keeffe’s three-letter RCSR codes \cite{okeef08}. Due to Jahn-Teller distortion, magnetic interactions between Cu$^{2+}$ ions ($J_{1}, J_{2}, J_{3}$, and $J_{4}$) are anisotropic (see the main text). The inset shows a photo image of a single crystal of \cuox.}
\end{figure}

\cuox\ is a promising material for hosting an exotic spin-liquid ground state. This Cu oxalate framework compound crystallizes in a monoclinic structure with space group $P2_{1}/c$ \cite{zhang12a}. The Cu ions form a hyper-honeycomb structure, also known as the (10.3)b or {\bf ths} net, connected by oxalate linkers, as shown in Fig.~\ref{fig:network} \cite{momma11}. The Kitaev model has been exactly solved on the (10.3)b net, and has a quantum spin-liquid ground state with gapless Majorana modes and gapped vison excitations arising from the emergent $\mathbb{Z}_2$ gauge field \cite{Trebst-class}. However, in \cuox\ there are two distinct Cu$^{2+}$ $S=1/2$ ions; the  Cu1 ions form chains along the $c$ axis [Fig.~\ref{fig:network}(a)] while the Cu2 ions form chains along the $a$ axis [Fig.~\ref{fig:network}(b)]. Due to Jahn-Teller distortion, the magnetic interactions between the Cu ions are anisotropic. Based on the orbital analysis and the distance between the Cu sites \cite{zhang18}, the magnetic interaction between the Cu2 ions with the shorter interatomic distance, $J_{4}$, is expected to be a strong antiferromagnetic (AF) interaction ($J_{4}/\mathrm{k_{B}}$> 200 K). The interaction between Cu1 ions, $J_{1}$, is expected to be smaller and also AF ($J_{1}/\mathrm{k_{B}}\sim$ 60 K). The interaction between Cu1 and Cu2 ions, $J_{2}$, is expected to be weakly AF, with magnitude $J_{2}/\mathrm{k_{B}}\sim$ 25 K. The interaction  $J_{3}$ between Cu2 ions with the larger interatomic distance is much weaker than the other interactions. 

The ground state of \cuox\ is quite exotic. The material is experimentally confirmed to be a magnetic insulator, whereas  density functional theory calculations find a metal \cite{jacko21}. This discrepancy suggests that the electronic correlations between electrons on the Cu$^{2+}$ ions lead to a Mott insulating phase and play an important role in the low-energy physics of this material \cite{jacko21}. Muon spin rotation ($\mu$SR) \cite{zhang18} and nuclear magnetic resonance (NMR) \cite{ding22} detect no magnetic order down to 50 mK. Despite the Mott insulating state, low-temperature ($T$) specific heat $C_{\mathrm{p}}$ down to 2 K reveals a large linear-in-$T$ contribution, suggestive of charge-neutral fermionic excitations in this compound \cite{zhang18}. To explain this exotic ground state, Zhang {\it et al.} proposed that a quantum spin liquid with a resonating-valence-bond (RVB) state may be realized in this system \cite{zhang18}. However, it was later argued, on the basis of a combined density-functional theory (DFT) and quantum many-body theory \cite{jacko21}, that the quantum disorder observed in \cuox\ arises because the Cu2 ions form dimers along the $J_4$ bonds [Fig.~\ref{fig:network}(c)], leaving weakly coupled chains of Cu1 ions along the $J_1$ bonds. Indeed it was proposed \cite{jacko21} that the effective interchain coupling should be unfrustrated but extremely weak, $\sim J_2^2/2J_4$, leading to an extremely low N\'eel temperature even though the system is bipartite.   

In this paper, we probe the low-energy physics of single crystals of the MOF oxalate \cuox\ through a systematic study of specific heat $C_{\mathrm{p}}$ in applied magnetic fields $H$.  We confirm the absence of magnetic order down to 180 mK in $H$ = 0 and observe a sizable zero-field linear-in-$T$ contribution to $C_{\mathrm{p}}$ at low $T$. Applied magnetic fields strongly suppress the low-temperature specific heat, with characteristic gapped behavior. We calculate the four-component relativistic band structure and thence estimate the Dzyaloshinskii–Moriya interaction (DMI). We find that the component of the DMI perpendicular to the $b$ axis alternates along the $J_1$ bonds. The low-energy physics of a quasi-one-dimensional antiferromagnetic Heisenberg chain (AFHC) in a magnetic field with an alternating DMI perpendicular to the field is described by the sine-Gordon model \cite{oshik97}. We show that the measured field dependence and magnitude of the gap are in good agreement with those expected for solitons, calculated from the sine-Gordon model using the parameters estimated from our DFT calculations. We also find that the $T\rightarrow 0$ K Wilson ratio estimated from our measurements is $R_{\mathrm{W}}=2.1(1)$, which agrees with the value $R_{\mathrm{W}}=2$ expected for spinons in spin-1/2 AFHCs \cite{johns00}.
This strongly suggests that \cuox\ is a quasi-one-dimensional Heisenberg antiferromagnet and not an RVB quantum spin liquid.

\begin{figure}[thb]
\includegraphics[width=0.95\linewidth]{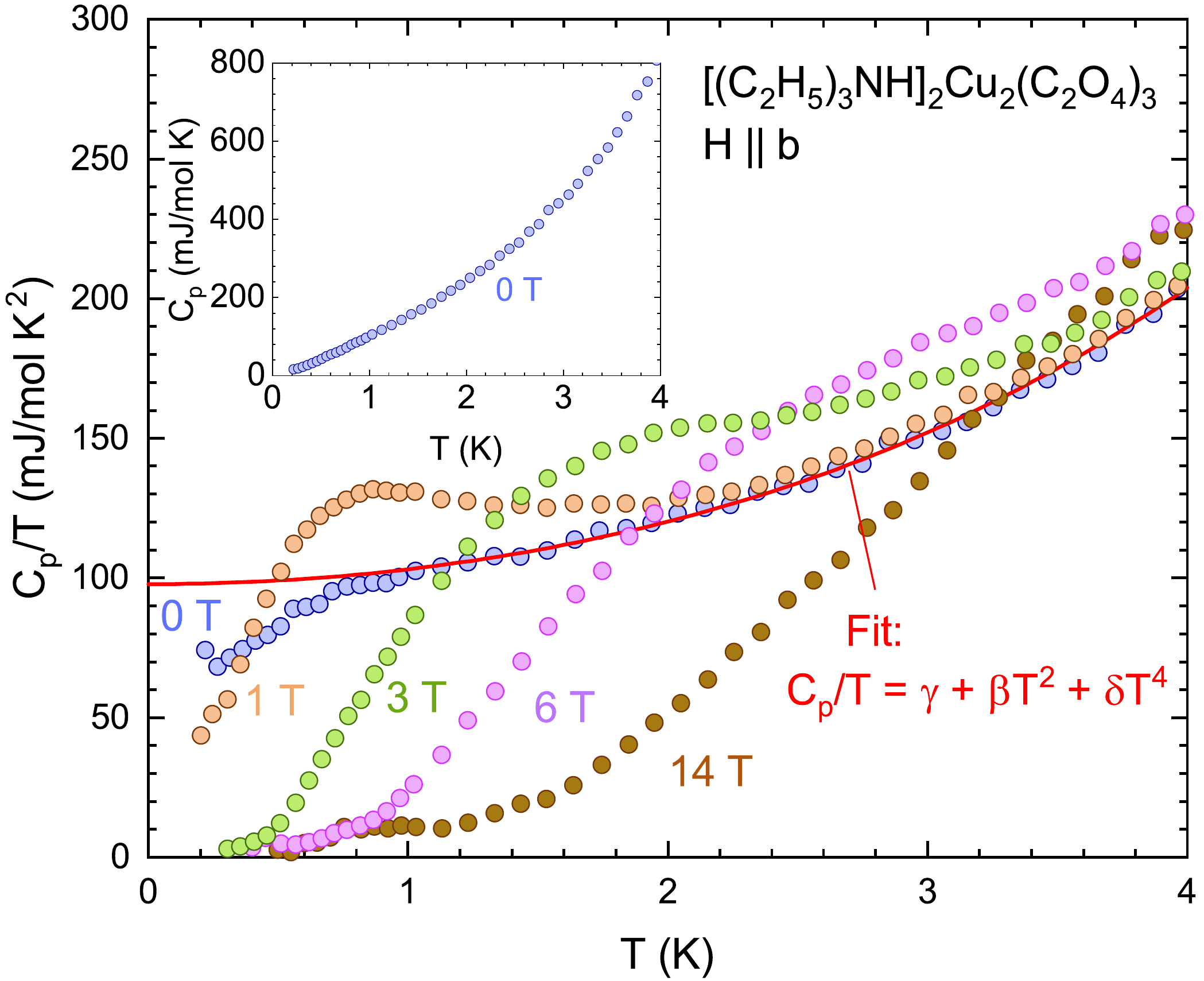}
\caption{\label{fig:ConTvsT} Specific heat $C_{\mathrm{p}}$ divided by temperature, $C_{\mathrm{p}}/T$, of \cuox\ as a function of $T$. Magnetic fields are applied along the $b$ axis. The red solid line is a fit to the data for $H$ = 0 T from 1 to 5 K using $C_{\mathrm{p}}/T = \gamma + \beta T^{2} + \delta T^{4}$. The $\gamma$ value is obtained to be 98(1) mJ/mol K$^{2}$. Inset: Temperature dependence of $C_{\mathrm{p}}$ at $H=$ 0 T. Below about 2 K, $C_{\mathrm{p}}(H=0~\mathrm{T})$ is approximately proportional to $T$.}
\end{figure}

\section{Experimental details}
Single crystals of \cuox\ were synthesized by an aqueous-solution method similar to that reported earlier \cite{zhang18}. Blue diamond-shaped crystals were extracted by vacuum filtration [Fig.~\ref{fig:network}(c) inset]. We used several crystals aligned along the $b$ axis. Specific-heat measurements were conducted using the relaxation method with a homemade calorimeter. The magnetic susceptibility $\chi$ was measured using a magnetic property measurement system from Quantum Design, Inc. (1 T = 10$^{4}$ Oe). Torque magnetometry was performed using a capacitive cantilever in a 35 T resistive magnet at the National High Magnetic Field Laboratory, Tallahassee, Florida.

\section{Results and discussion}
The specific heat, $C_{\mathrm{p}}(T)$, of \cuox\ shown in Fig.~\ref{fig:ConTvsT} exhibits an unusual $T$ dependence at low $T$. We observe no magnetic order in $C_{\mathrm{p}}$ down to $T =$ 180 mK, consistent with the previous reports from $\mu$SR \cite{zhang18} and NMR \cite{ding22}, as shown in the inset to Fig.~\ref{fig:ConTvsT}. The zero-field specific heat is approximately proportional to $T$ at low $T$, indicative of the presence of gapless charge-neutral magnetic excitations in this Mott insulator.

Fitting $C_{\mathrm{p}}/T = \gamma  +\beta T^{2}  +\delta T^{4}$ to the data for $H$ = 0~T from 1 to 5 K (data are presented only below 4 K in Fig.~\ref{fig:ConTvsT}), we extract the linear-in-$T$ contribution to the zero-field specific heat, $\gamma T$, associated with gapless charge-neutral magnetic excitations. The obtained value of $\gamma$ per mole of formula units is 98(1) mJ/mol K$^{2}$, somewhat larger than that reported in the previous work ($\gamma=36$ mJ/Cu-mol K$^{2}$ = 72 mJ/mol K$^{2}$) \cite{zhang18}. We observe a deviation from the fitted curve below $T$ = 0.7 K. We will discuss this deviation later.

\begin{figure}[t]
\includegraphics[width=0.95\linewidth]{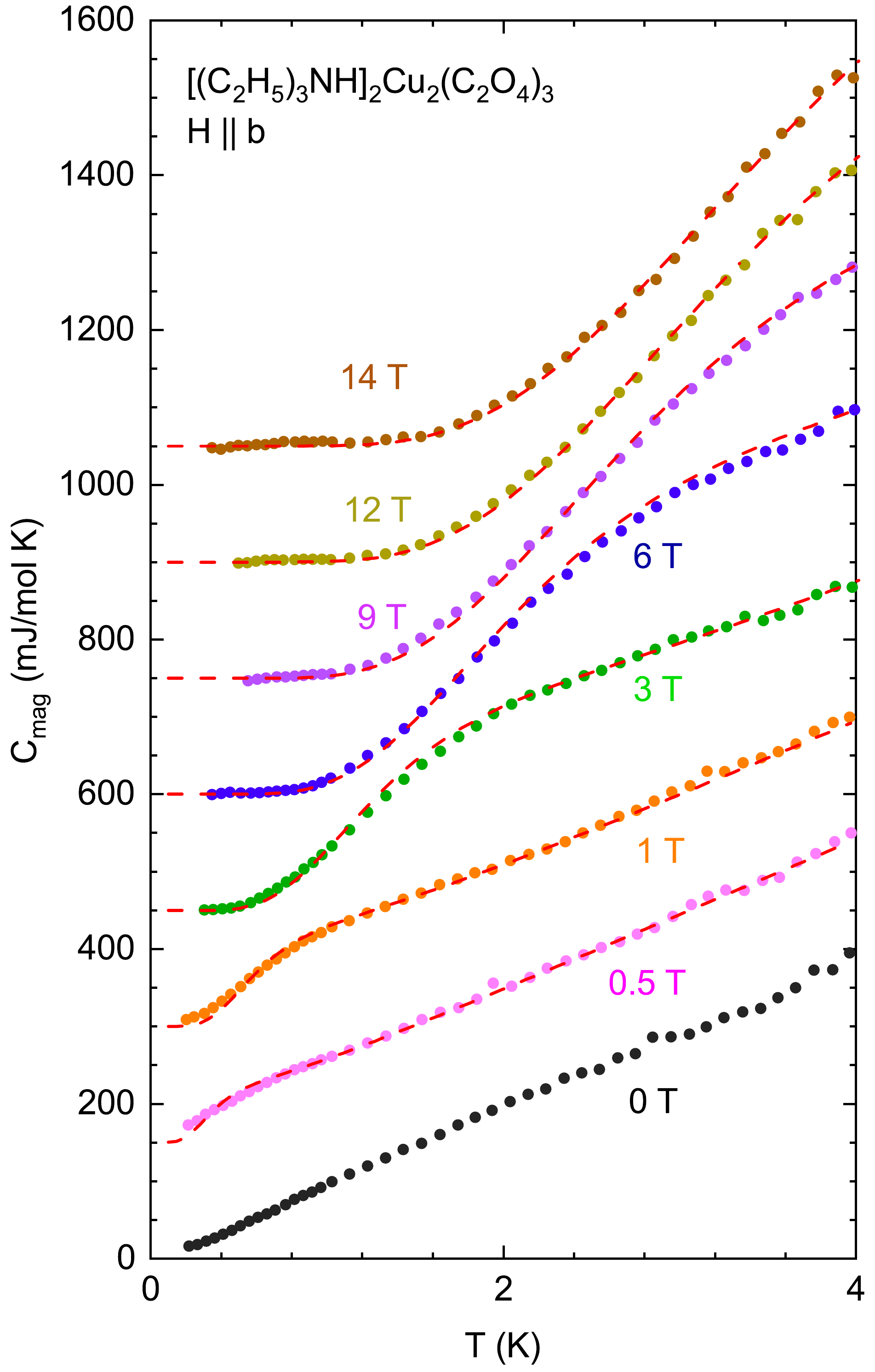}
\caption{\label{fig:cm} Temperature dependence of the magnetic contribution to the specific heat of \cuox\, where $C_{\mathrm{mag}}=C_{\mathrm{p}}-C_{\mathrm{ph}}$ and $C_{\mathrm{p}}$ is the phonon contribution. The curves are successively shifted upward by 150 mJ/mol K for clarity. The red dashed lines are fits to the sine-Gordon model (see the main text). A clear development of an energy gap is observed upon applying magnetic fields parallel to the $b$ axis.}
\end{figure}

As shown in Fig.~\ref{fig:ConTvsT}, the large linear-in-$T$ specific heat term due to gapless magnetic excitations gradually diminishes with applied magnetic field parallel to the $b$ axis, which is perpendicular to the quasi-one-dimensional chains formed by Cu1 (along the $c$ axis) and Cu2 (along the $a$ axis) ions. At $H$ = 1 T, with decreasing temperature, $C_{\mathrm{p}}/T$ exhibits a broad peak at $T\approx$ 0.8~K in Fig.~\ref{fig:cm} and then rapidly decreases. At larger magnetic fields, the peak shifts to higher temperatures, and $C_{\mathrm{p}}/T$ becomes zero at low temperatures, suggestive of the formation of an $H$-induced gap for spin excitations. The gap can be seen more clearly in the temperature dependence of the magnetic specific heat $C_{\mathrm{mag}}$ discussed below.

\begin{figure}[t]
\includegraphics[width=0.95\linewidth]{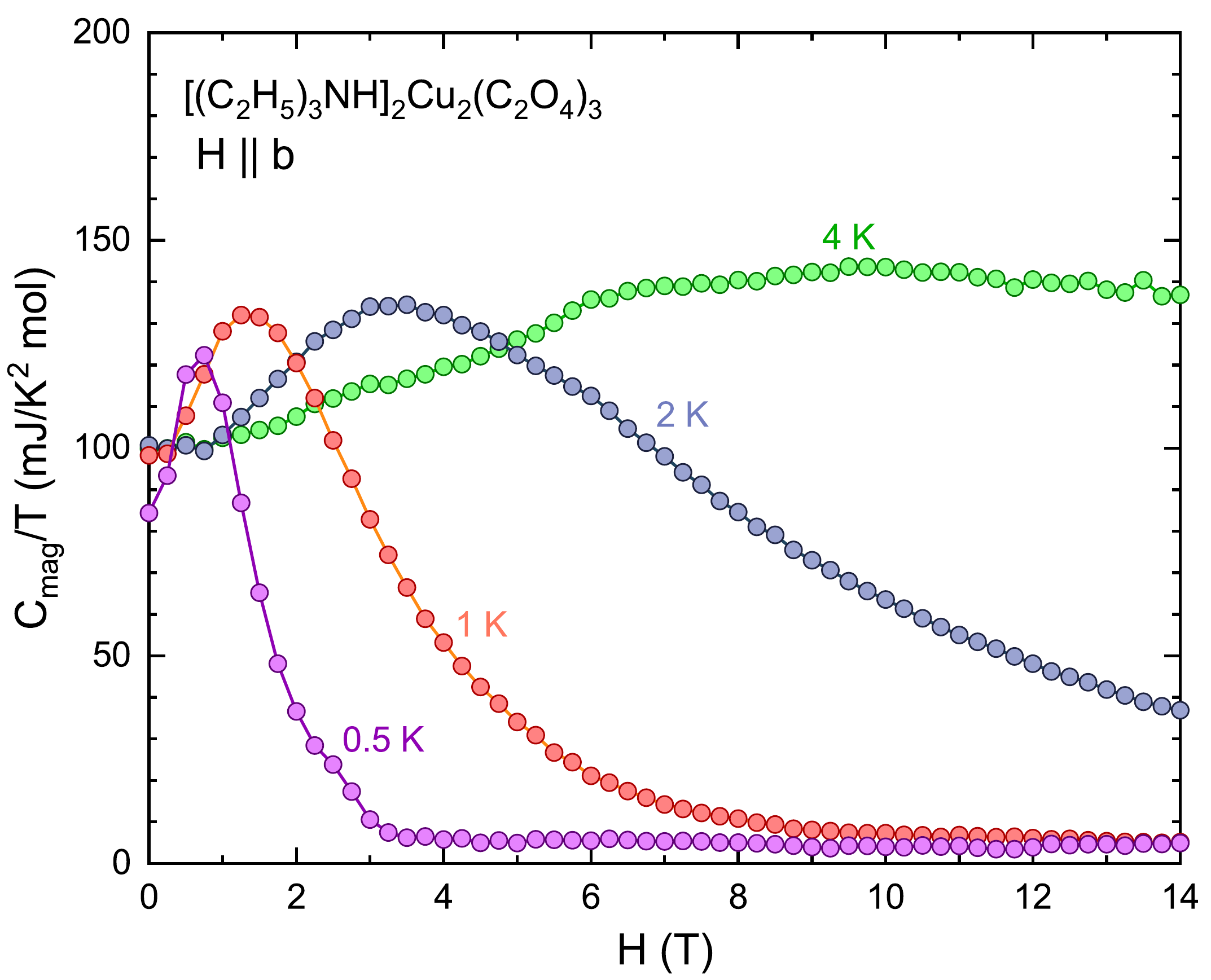}
\caption{\label{fig:CmTvsH} Magnetic field $H$ dependence of the magnetic contribution to the specific heat $C_{\mathrm{mag}}(T)/T$ of \cuox at four temperatures. The magnetic fields are applied along the $b$ axis. The peak position in $C_{\mathrm{mag}}(T)/T$ shifts to a higher magnetic field with increasing temperature.}
\end{figure}

To investigate the magnetic contribution to the specific heat, we subtract the phonon contribution, $C_{\mathrm{ph}} = \beta T^{3}  +\delta T^{5}$, from the total specific heat $C_{\mathrm{p}}$, and plot $C_{\mathrm{mag}}(T) = C_{\mathrm{p}}(T) - C_{\mathrm{ph}}(T)$ in Fig.~\ref{fig:cm} for each $H$. We clearly observe a crossover from linear-in-$T$ to gapped behavior with increasing magnetic fields. The field-induced gap is also detected in the field dependence of $C_{\mathrm{mag}}/T$, as shown in Fig.~\ref{fig:CmTvsH}. At $T=$ 0.5 K, $C_{\mathrm{mag}}/T$ due to gapless excitations at zero field decreases rapidly with magnetic field, after showing a peak at about 0.5 T. Above 3~T, $C_{\mathrm{mag}}(H)/T$ is nearly independent of magnetic field at $T=0.5$ K, and its magnitude is about 5\% of $ C_{\mathrm{mag}}(H=0~\mathrm{T})/T$, indicating that the ground state is completely gapped out in high magnetic fields. Upon increasing the temperature, the peak moves to higher magnetic fields and becomes broader.

The absence of magnetic order and the large linear-in-T contribution to $C_{\mathrm{p}}$ at low temperatures observed in zero magnetic field corroborate a charge-neutral gapless ground state, such as a quantum spin-liquid state. To explain these observations, Zhang {\it et al.} proposed a resonating-valence-bond (RVB) state in a hyper-honeycomb lattice of Cu$^{2+}$ ions associated with reduced dimensionality \cite{zhang18}. The proposed RVB state, however, appears to be unlikely because in a spin-1/2 hyper-honeycomb lattice without strong spin-orbit coupling, magnetic frustrations, which are key ingredients to such spin-liquid states, are absent.

Instead, we attribute this behavior observed in \cuox\ to weakly coupled one-dimensional antiferromagnetic Heisenberg chains (AFHCs) stemming from a distorted hyper-honeycomb lattice elongated in the $b$-axis direction due to the Jahn-Teller distortion of the CuO$_{6}$ octahedra. Jacko and Powell have pointed out that the Jahn-Teller distortion can cause the formation of quasi-one-dimensional spin chains of Cu1 ions along the $c$ axis and dimerization of Cu2 ions along the $a$ axis \cite{jacko21}. The  values of the exchange interactions extracted from the DFT and perturbation theory depend somewhat on the assumed values of the on-site Coulomb (Hubbard) interaction, $U$, and Hund's rule coupling, $J_{\mathrm{H}}$. However, the relative magnitudes of the interactions are $J_4 \gg J_1 \gg J_2 \gg |J_3|$ for reasonable choices of $U$ and $J_{\mathrm{H}}$, and the numerical values are consistent with those estimated in Ref. [\onlinecite{zhang18}] from orbital analysis. Given that $J_{3}$ is negligible compared with $J_4$, the interchain coupling strength $J_{\perp}$ can be estimated to be $J_{\perp}/\mathrm{k_{B}}=J_{2}^{2}/(2J_{4}\mathrm{k_{B}})\simeq$ 0.1 K, suggesting that \cuox\ can be treated as an AFHC.

In this situation, the zero-field linear-in-$T$ contribution, $\gamma T$, to the specific heat can be attributed to the spinon contribution in one-dimensional spin-1/2 AFHCs. The spinon specific-heat coefficient is given by $\gamma=\frac{2R\mathrm{k_{B}}}{3J}$, where $R$ is the molar gas constant and $J$ the exchange interaction between spins \cite{johns00}. Extracted from the measured $\gamma$ in the present work, $J/\mathrm{k_{B}}$ is estimated to be 56.7(8) K, in excellent agreement with $J_{1}$ between Cu1 ions estimated from an orbital analysis in Ref. \cite{zhang18}.

\begin{figure}[t]
\includegraphics[width=0.95\linewidth]{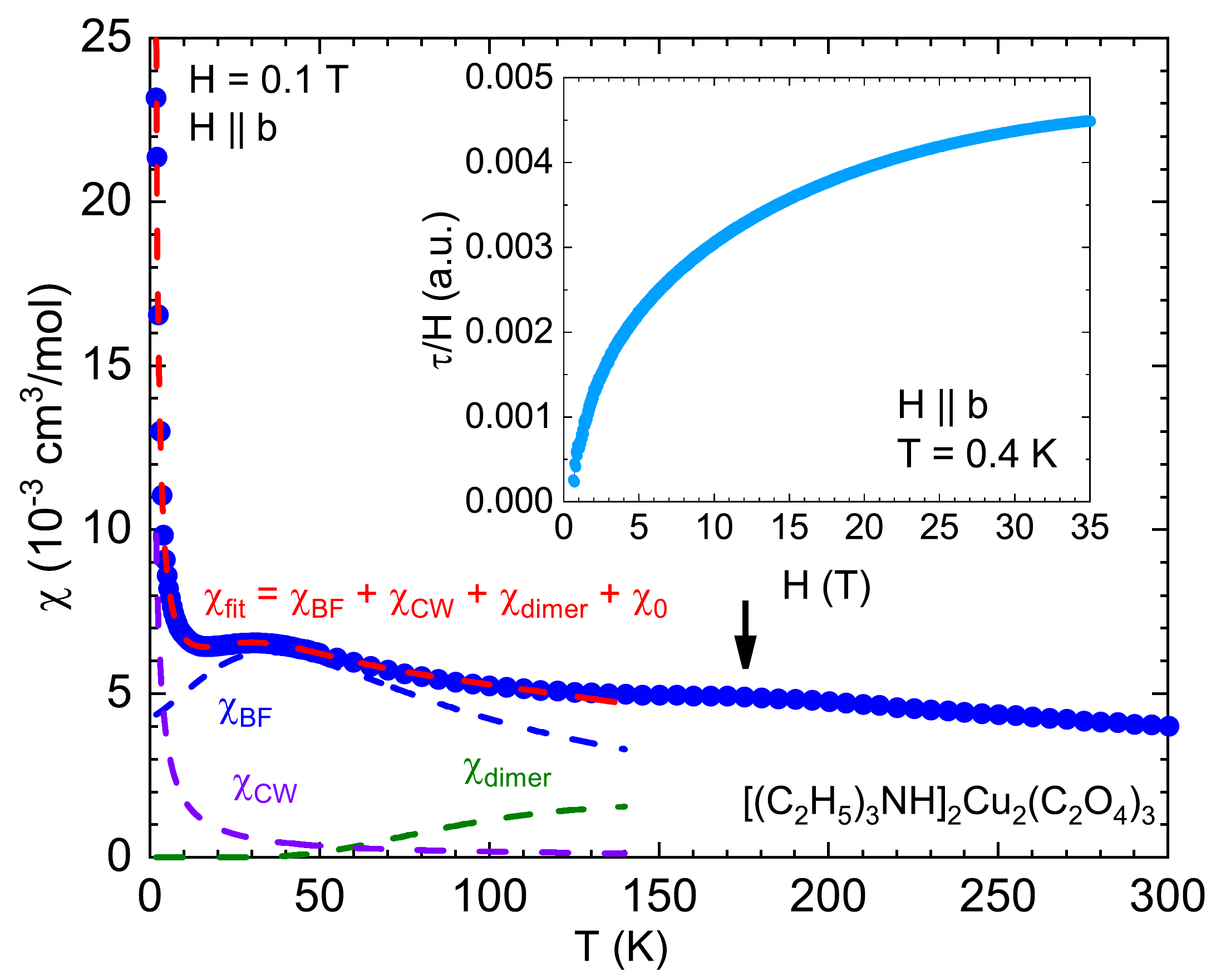}
\caption{\label{fig:torque} Temperature dependence of magnetic susceptibility $\chi$ in magnetic field $H$ = 0.1 T. The magnetic field is applied parallel to the $b$ axis. The red line is a fit to the data from 2 to 140 K using $\chi_{\mathrm{fit}} = \chi_{\mathrm{BF}}+\chi_{\mathrm{CW}}+\chi_{\mathrm{dimer}}+\chi_{0}$, where $\chi_{\mathrm{BF}}$ is the magnetic susceptibility for one-dimensional AFHCs (the blue dashed line), $\chi_{\mathrm{CW}}$ is a contribution from impurities/defects (the purple dashed line), $\chi_{\mathrm{dimer}}$ is a contribution from AFHDs (the green dashed line), and $\chi_{0}$ is a diamagnetic susceptibility estimated from the Pascal's constants \cite{bain08,zhang12a}. The arrow indicates an anomaly associated with an order-disorder transition of (C$_{2}$H$_{5}$)$_{3}$NH$^{+}$ cations at $\sim$170 K \cite{zhang12a}. Inset: Field dependence of magnetic torque divided by magnetic field, $\tau/H = M_{\perp}V$, where $M_{\perp}$ is the magnetization perpendicular to applied magnetic field and $V$ is the sample volume, of \cuox\ at $T = 0.4$ K. The magnetic field is applied along the $b$ axis. No discernible anomaly associated with a magnetic transition is observed.}
\end{figure}

The magnetic susceptibility $\chi(T)$ of this material also supports this scenario. The overall $T$ dependence of $\chi$ at $H =$ 0.1 T ($H\parallel b$) is plotted in Fig.~\ref{fig:torque}. Associated with an order-disorder transition of (C$_{2}$H$_{5}$)$_{3}$NH$^{+}$ cations \cite{zhang18}, an anomaly indicated by the black arrow is observed at $T\sim$ 170 K. Upon decreasing temperature, $\chi$ exhibits a broad maximum at $T_{\mathrm{max}} \sim$ 35 K, indicating the presence of an AF exchange interaction between the Cu ions in one-dimensional chains. At low $T$, $\chi$ shows a Curie-Weiss-type divergence, most likely due to impurity and/or defect spins. This $T$ dependence of $\chi$ in \cuox\ is well described by the sum of the contributions from one-dimensional AFHCs, antiferromagnetic Heisenberg dimers (AFHDs), impurity/defect spins, and a diamagnetic term. We analyze $\chi(T)$ quantitatively by using the following equations,
\begin{align}
  \chi_{\mathrm{fit}}& = \chi_{\mathrm{BF}} + \chi_{\mathrm{dimer}} + \chi_{\mathrm{CW}}  + \chi_{0},\\
    \chi_{\mathrm{BF}}& \nonumber\\
  =&\frac{N_{\mathrm{A}}g^{2}\mu_{\mathrm{B}}^{2}}{\mathrm{k_{B}}T}\frac{0.25+0.14995x+0.30094x^{2}}{1+1.9862x+0.68854x^{2}+6.0626x^{3}},\\
  \chi_{\mathrm{dimer}}& =\frac{N_{\mathrm{A}}g^{2}\mu_{\mathrm{B}}^{2}}{J_{\mathrm{dimer}}}\frac{e^{-1/t}}{t}\frac{1}{1+3e^{-1/t}},\\
  \chi_{\mathrm{CW}}&=\frac{C_{\mathrm{CW}}}{T-\theta},
\end{align}
where $\chi_{\mathrm{BF}}$ is the Bonner–Fisher result \cite{bonne64}, parameterized with $x=J/(2\mathrm{k_{B}}T)$ by \cite{estes78}, $N_{\mathrm{A}}$ is Avogadro’s number, $g$ is the spectroscopic-splitting factor, $\mu_{\mathrm{B}}$ is the Bohr magneton, $\theta$ is the Weiss temperature, $\chi_{\mathrm{CW}}$ is a Curie-Weiss contribution, $\chi_{\mathrm{dimer}}$ is the magnetic susceptibility of the AFHDs \cite{bonne79,johns00}, $J_{\mathrm{dimer}}$ is the exchange interaction between dimerized Cu$^{2+}$ ions, $t=J_{\mathrm{dimer}}/\mathrm{k_{B}}T$, and $\chi_{0}$ is the diamagnetic susceptibility ($-2.77\times 10^{-4}$ cm$^{3}$/mol) estimated from Pascal's constants \cite{bain08,zhang12a}. As shown in Fig.~\ref{fig:torque}, the fit is in excellent agreement with our data from 2 to 140 K. The obtained fitting parameters are $J/\mathrm{k_{B}} = 54(1)$~K,  $g=2.48(3)$, $C_{\mathrm{CW}}=1.78(2)\times 10^{-2}$~cm$^{3}$K/mol, $\theta=0.98(1)$~K, and $J_{\mathrm{dimer}}=280(10)$ K. The obtained $J=$ 54(1)~K agrees with that found from our specific-heat measurements. The obtained $J_{\mathrm{dimer}}$ is also consistent with $J_{4}$ from theoretical calculations \cite{zhang18,jacko21}. Combining the obtained $g$ factor, $\chi_{\mathrm{BF}}(T\rightarrow 0~\mathrm{K})$, and $\gamma$ in the present work, we evaluate the Wilson ratio,
\begin{align}
R_{\mathrm{W}}=\frac{4}{3}\left (\frac{\pi \mathrm{k_{B}}}{g\mu_{\mathrm{B}}}\right )^{2}\frac{\chi}{\gamma}.
\end{align}
We obtain $R_{\mathrm{W}}=2.1(1)$, which agrees with the expected value of $R_{\mathrm{W}}=2$ for one-dimensional spin-1/2 AFHCs \cite{johns00}.

However, the one-dimensional spin-1/2 AFHC remains gapless in magnetic fields up to the saturation field. Our magnetic-torque data demonstrate no magnetic transition nor saturation up to 35 T, as shown in the inset to Fig.~\ref{fig:torque}. 
To better understand this, we performed four-component relativistic DFT calculations of the electronic structure  in an all-electron full-potential local orbital basis code (FPLO) \cite{koepernik99}, using the Perdew-Burke-Ernzerhof generalized gradient approximation \cite{perdew96}. The density was converged on an $(8 \times 8 \times 8)$ $k$ mesh. FPLO implements the full-potential non-orthogonal local-orbital minimum basis band-structure scheme \cite{koepernik99}. All calculations were performed for the measured crystal structure \cite{zhang18}. 
These calculations are thus the four-component analogue of the scalar relativistic calculations reported in Ref. \cite{jacko21}. 

\begin{table*}[t]
	\centering
	\begin{tabular}{ccccccc} 
		\hline
		$n$ & Cu1 & Cu2 & 1 &   2 & 3 & 4  \\  
		\hline 
		$t_{n}$ &
			 $\left(\begin{array}{cc} -376 & -149 \\ -149 & -271 \end{array}\right)$ & 
			 $\left(\begin{array}{cc} -417 & -94 \\ -94 & -237 \end{array}\right)$ & 
			 $\left(\begin{array}{cc} -29 & -90 \\ 119 & 169 \end{array}\right)$ & 
			 $\left(\begin{array}{cc} 245 & -37 \\ -44 & -36 \end{array}\right)$ & 
			 $\left(\begin{array}{cc} 36 & 124 \\ 124 & 141 \end{array}\right)$ & 
			 $\left(\begin{array}{cc} 36 & -102 \\ -102 & 199 \end{array}\right)$ \\
		$\lambda_{n}^{a*}$ &  
			$\left(\begin{array}{cc} 0 & \pm7 \\ \mp7 & 0 \end{array}\right)$ & 
			$\left(\begin{array}{cc} 0 & 7 \\ -7 & 0 \end{array}\right)$ & 
			$\left(\begin{array}{cc} \mp2 & \mp12 \\ \mp4 & \pm9 \end{array}\right)$ & 
			$\left(\begin{array}{cc} 2 & 18 \\ -16 & -2 \end{array}\right)$ & 
			$\left(\begin{array}{cc} 0 & -9 \\ 9 & 0 \end{array}\right)$ & 
			$\left(\begin{array}{cc} 0 & -8 \\ 8 & 0 \end{array}\right)$ \\
		$\lambda_{n}^{b}$ &  
			$\left(\begin{array}{cc} 0 & -25 \\ 25 & 0 \end{array}\right)$ & 
			$\left(\begin{array}{cc} 0 & -12 \\ 12 & 0 \end{array}\right)$ & 
			$\left(\begin{array}{cc} -3 & -15 \\ -14 & 4 \end{array}\right)$ & 
			$\left(\begin{array}{cc} -4 & 2 \\ 1 & 2 \end{array}\right)$ & 
			$\left(\begin{array}{cc} 0 & 14 \\ -14 & 0 \end{array}\right)$ & 
			$\left(\begin{array}{cc} 0 & -12 \\ 12 & 0 \end{array}\right)$ \\
		$\lambda_{n}^{c}$ &  
			$\left(\begin{array}{cc} 0 & \pm13 \\ \mp13 & 0 \end{array}\right)$ & 
			$\left(\begin{array}{cc} 0 & 5 \\ -5 & 0 \end{array}\right)$ & 
			$\left(\begin{array}{cc} \pm1 & \pm9 \\ \pm9 & \pm9 \end{array}\right)$ & 
			$\left(\begin{array}{cc} -3 & -10 \\ 17 & -2 \end{array}\right)$ & 
			$\left(\begin{array}{cc} 0 & 3 \\ -3 & 0 \end{array}\right)$ & 
			$\left(\begin{array}{cc} 0 & 9 \\ -9 & 0 \end{array}\right)$ \\
		\hline 
	\end{tabular}
	\caption{Tight-binding parameters calculated from DFT. Here $n=\text{Cu1}$ or Cu2 indicates on-site parameters for the given sublattice, and $n=1-4$ indicates a bond labeled in Fig.~\ref{fig:network}(c) by the nearest-neighbor interaction $J_{n}$. The symbols $\pm$ and $\mp$ in the Cu1 ($n=1$) columns indicate that the spin-orbit-coupling parameter alternates on neighboring ions (bonds). All values are in units of meV.}
	\label{tab:tb}
      \end{table*}
      
As expected for the light elements in \cuox, we find only minor difference in the band structure calculated from the four-component and scalar relativistic theories. From the overlaps of the Wannier orbitals extracted from the four-component calculation, we obtain a tight-binding model of the form \cite{Jacko17,Khosla}
\begin{align}
	\hat{{\cal H}}_{\mathrm{tb}}
	=&\sum_{i,j} \hat{\bm{d}}_{i\alpha}^\dagger \left( t_{ij}\delta_{\alpha\beta} + i\bm{\lambda}_{ij}\cdot\bm{\sigma}_{\alpha\beta} \right) \hat{\bm{d}}_{j\beta},
\end{align}
where $\hat{\bm{d}}_{i\alpha}^\dagger=(\hat{d}_{i1\alpha}^{\dagger},\hat{d}_{i2\alpha}^{\dagger})$, $\hat{d}_{i\mu\alpha}^{\dagger}$ $(\hat{d}_{i\mu\alpha})$ creates (annihilates) an electron with spin $\alpha$ in the $\mu$th orbital on the $i$th Cu ion, $\bm{\sigma}$ is the vector of Pauli matrices, $t_{ij}$ is the scalar hopping integral, and $\bm{\lambda}_{ij}$ is the vectorial spin-orbit-coupling hopping integral. The values of the hopping integrals calculated from DFT are given in Table~\ref{tab:tb}.

We model the electronic interactions on each atom by an $e_g$-symmetry Kanamori interaction \cite{Kanamori,Georges}
\begin{eqnarray}
	\hat{V}_K^{e_g} &=& 
	U \sum_{i,m} \hat{n}_{im\uparrow} \hat{n}_{im\downarrow} + U' \sum_{i,m \neq m'}\hat{n}_{im \uparrow} \hat{n}_{im' \downarrow} \nonumber \\
	&&+ (U'-J_{\mathrm{H}}) \sum_{i,m < m',\sigma}\hat{n}_{im \sigma} \hat{n}_{im' \sigma} \nonumber \\
	&&- J_{\mathrm{H}} \sum_{i,m \neq m'} \hat{d}_{im\uparrow}^\dagger \hat{d}_{im\downarrow} \hat{d}_{im'\downarrow}^\dagger \hat{d}_{im'\uparrow} \nonumber \\
	&&+ J_{\mathrm{H}} \sum_{i,m \neq m'} \hat{d}_{im\uparrow}^\dagger \hat{d}_{im\downarrow}^\dagger \hat{d}_{im'\downarrow} \hat{d}_{im'\uparrow},
\end{eqnarray}
where $U$ ($U'$) is the intra-(inter-)orbital Coulomb repulsion,  $J_{\mathrm{H}}$ is the Hund's rule coupling, and $\hat{n}_{im \sigma}=\hat{d}_{im \sigma}^\dagger\hat{d}_{im \sigma}$. 
For simplicity, we approximate the on-site inter-orbital Coulomb repulsion as $U' = U-2J_{\mathrm{H}}$, which is exact in octahedral symmetry \cite{Georges}.

We calculate the magnetic interactions by perturbing around the large $U$ limit to second order in $t_{ij}$ and first order in $\bm{\lambda}_{ij}$ \cite{Powell,Jacko20}.
The values of $U$ and $J_{\mathrm{H}}$ are not accurately known \textit{a priori} and have limited transferability between materials. Therefore, we estimate them by comparing the calculated value of $J_1$ to that obtained experimentally above. For example, for $U=6$~eV and $J_{\mathrm{H}}=0.3$~eV we find $J_1=54$~K, consistent with the experimental estimate; therefore we fix $U$ and $J_{\mathrm{H}}$ at these values.

Since we include spin-orbit coupling, we also find Dzyaloshinskii-Moriya interactions (DMIs), Table~\ref{tab:DM}. Below we focus on the DMI along the $J_1$ bonds, $\bm{D}_1=(D_1^{a^*}, D_1^b, D_1^c)$, where the $a^{*}$ axis is perpendicular to both the $b$ and $c$ axes and $D_1^n$ is the component of the DMI along the $n$ axis, because this is crucial for understanding the experiments described above. We find, independent of the values of $U$ and $J_{\mathrm{H}}$, that $D_1^b$ is the same for all bonds, whereas $D_1^{a^*}$ and $D_1^c$ alternate in sign from bond to bond. Thus the Hamiltonian for a single chain is
\begin{align}
	\hat{{\cal H}}_1 =\sum_i  \Big\{ J_1  \hat{\bm{S}}_i \cdot \hat{\bm{S}}_{i+1} + \left[ \bm{D}_1^\| + (-1)^i \bm{D}_1^\perp \right] \cdot \hat{\bm{S}}_i \times \hat{\bm{S}}_{i+1} \notag\\ +g\mu_{\mathrm{B}}H \hat{S}_i^b \Big\},
\end{align}
where $\bm{D}_1^\|=(0, D_1^b, 0)$ is the (non-alternating) component of the DMI parallel to the $b$ axis and $\bm{D}_1^\perp=(D_1^{a^*}, 0, D_1^c)$ is the (alternating) component of the DMI perpendicular to the $b$ axis.

\begin{table}[tb]
\centering
\begin{tabular}{ccccc} 
	\hline
	$n$ & 1 &  2 & 3 & 4  \\  
	\hline 
	$J_n$ & 54 & 21 & 3 & 477 \\
	$\bm{D}_n=(D_n^{a^*}, D_n^b, D_n^c)$ & $(\pm 13, 3, \mp16)$ & $(1.1, 1.5, -0.6)$ & $\bm0$ & $\bm0$ \\
	\hline 
\end{tabular}
\caption{Calculated exchange interactions and DMIs for $U=6$~eV and $J_{\mathrm{H}}=0.3$~eV. All values are in units of K.}
\label{tab:DM}
\end{table}

In the presence of an external magnetic field, the Heisenberg chain with an alternating DMI perpendicular to the field can be mapped to an XXZ model in an alternating magnetic field \cite{oshik97} of magnitude
\begin{align}
	h = 
	H \sin\left[ \frac12\arctan\left( \frac{|\bm D_1^\perp|}{J_1} \right)\right].
\end{align}
This staggered field opens a gap, which can be described by the sine-Gordon model for one-dimensional spin-1/2 AFHCs \cite{essle98,oshik97}. In this model, the free energy of the system is expressed in terms of the solution of a single nonlinear integral equation for the complex quantity $\epsilon(\theta)$,
\begin{align}
  \epsilon(\theta)&= -i \Delta\beta\sinh(\theta +i\eta^{\prime}) \nonumber\\
                  &-\int_{-\infty}^{\infty}d\theta^{\prime}G_{0}(\theta-\theta^{\prime})\ln[1+\exp(-\epsilon(\theta^{\prime}))]\\
  &+\int_{-\infty}^{\infty}d\theta^{\prime}G_{0}(\theta-\theta^{\prime}+2i\eta^{\prime})\ln[1+\exp(-\bar{\epsilon}(\theta^{\prime}))]\nonumber,
\end{align}
where $\beta=1/\mathrm{k_{B}}T$, $\Delta$ is a soliton mass, or gap, $\bar{\epsilon}(\theta)$ is the complex conjugate of $\epsilon(\theta)$, and
\begin{align}
  G_{0}(\theta)=\int_{0}^{\infty}\frac{d\omega}{\pi^{2}}\frac{\cos(2\omega\theta/\pi)\sinh[\omega(\xi-1)]}{\sinh(\omega\xi)\cosh(\omega)}.
\end{align}
The free energy density is given by
\begin{align}
f(\beta)=-\frac{\Delta N_{\mathrm{A}}}{\pi v_{s}\beta}\textrm{Im}\int_{-\infty}^{\infty}d\theta 
\sinh(\theta+i\eta^{\prime})\ln[1+\exp(-\epsilon(\theta))],
\end{align}
where $v_{s}$ is the spinon velocity. The free energy density is independent of the value of $\eta^{\prime}$ if $0 < \eta^{\prime}<\pi\xi/2$. Assuming that the effect of the SU(2) symmetry breaking due to the applied magnetic field is negligible, we fix $\xi=1/3$ and $v_{s}=\pi J_{1}/2$. Utilizing $C_{\mathrm{p}}=-T\frac{\partial^{2}f(\beta)}{\partial T^{2}}$ and setting $\Delta$ to be a fitting parameter, we fit the specific heat obtained from the sine-Gordon model to the data in magnetic fields. The red dashed lines in Fig.~\ref{fig:cm} are the calculated specific heat curves, in excellent agreement with our data.

\begin{figure}[t]
\includegraphics[width=0.95\linewidth]{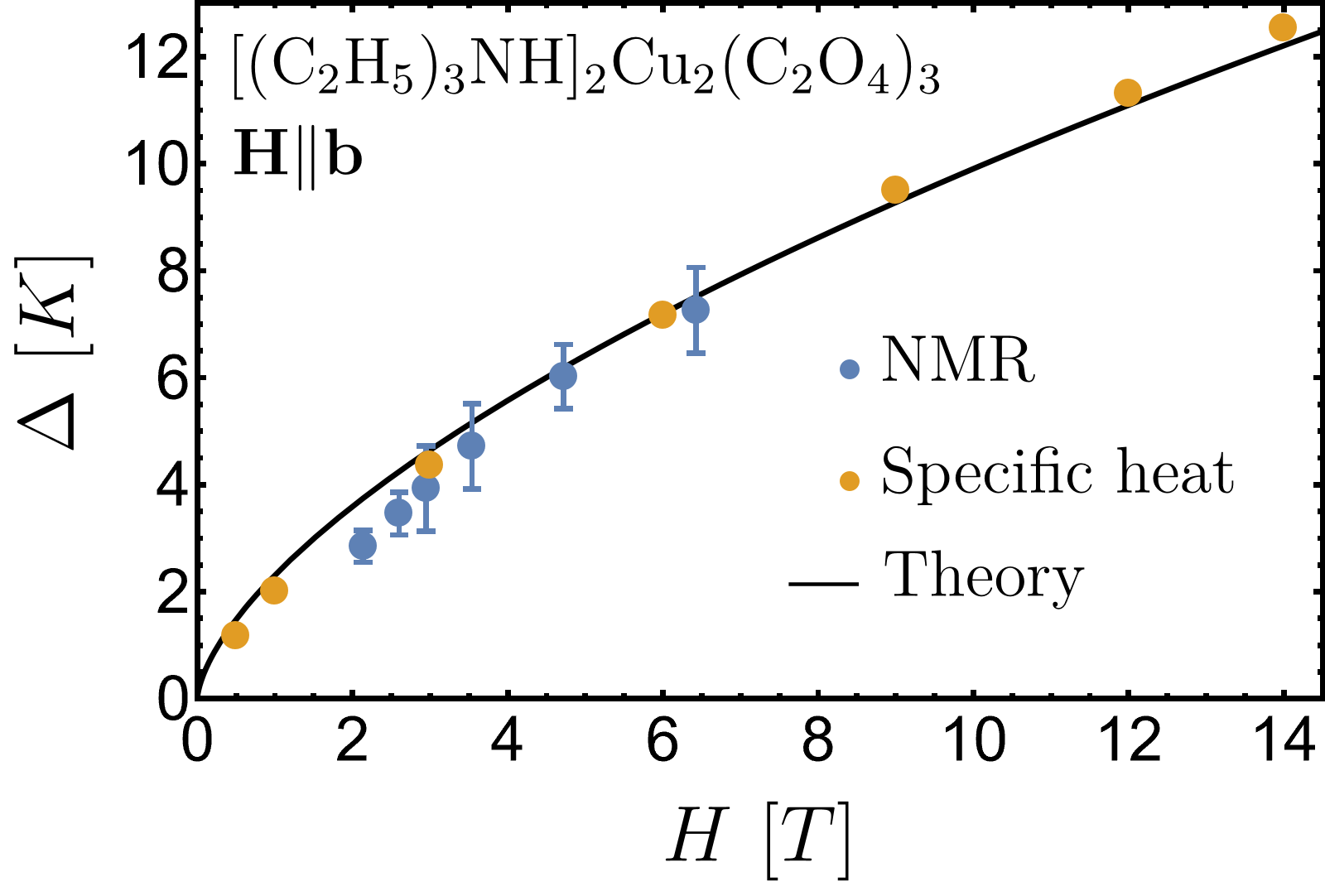}
\caption{\label{fig:delta} Field dependence of the soliton gap extracted from the magnetic specific heat and NMR data \cite{ding22}. The solid line is the soliton gap predicted \cite{oshik97} from the sine-Gordon model, Eq.~(\ref{eq:gap}), for the DMI and exchange interaction obtained from our DFT calculations. This curve contains no adjustable parameters once $U$ and $J_{\mathrm{H}}$ are fixed and depends only extremely weakly on the choice of these parameters.}
\end{figure}

The gap opened by the applied magnetic field is given by \cite{oshik97} 
\begin{align}
	\Delta(H) = 1.85 \left(\frac{g\mu_{\mathrm{B}}h}{2J_1}\right)^{2/3}J_1\left|\ln\left(\frac{g\mu_{\mathrm{B}}h}{2J_1}\right)\right|^{1/6}. \label{eq:gap}
\end{align}
The soliton gap extracted from the fits to the sine-Gordon model (Fig.~\ref{fig:cm}) is compared in Fig.~\ref{fig:delta} to this prediction. The agreement is excellent. Note that once $U$ and $J_{\mathrm{H}}$ have been fixed to reproduce the experimental estimate of $J_1$, there are no free parameters in Eq.~(\ref{eq:gap}). Furthermore, the curve predicted by Eq.~(\ref{eq:gap})  depends extremely weakly on the choice of $U$ and $J_{\mathrm{H}}$.
These results strongly indicate that the large linear-in-$T$ contribution to the specific heat and magnetic-field-induced gapped behavior stem from gapless spinons and gapped solitons in weakly coupled spin-1/2 one-dimensional AFHCs, rather than the RVB state proposed in Ref.~\cite{zhang18}.

Finally, we comment on the deviation from the fitting curve $C_{\mathrm{p}}/T=\gamma +\beta T^{2}+\delta T^{4}$ in the specific heat of \cuox\ in Fig.~\ref{fig:ConTvsT} at temperatures below 0.7 K $\sim 0.01J_1/\mathrm{k_{B}}$. In the similar temperature range below 1 K, NMR reveals a clear signature of slowing down of Cu$^{2+}$ spin dynamics \cite{ding22}. The deviation of the spinon contribution from linear-in-$T$ behavior in this temperature range may be related to this slow spin dynamics. We note that a similar deviation is also reported below $T=$ 1 K $\sim 0.01J/\mathrm{k_{B}}$ in the sine-Gordon spin system KCuGaF$_{6}$ \cite{umega12,umega15}. The resemblance between the two distinct materials may share a universal origin in one-dimensional spin-1/2 AFHCs, requiring further experimental investigations.

\section{Summary}
We have measured the specific heat of \cuox\ single crystals at very low temperatures in magnetic fields. We observe no anomaly in zero field and a sizable linear-in-$T$ contribution to the specific heat associated with spinon excitations. Upon applying magnetic fields, a soliton gap is induced, varying as $\Delta (H) \propto H^{2/3}$. These results can be explained by the sine-Gordon model for weakly coupled one-dimensional AFHCs as proposed in \cite{jacko21}, rather than an RVB state proposed in previous work \cite{zhang18}. Since \cuox\ undergoes no magnetic transition down to 50 mK and has weak interchain coupling, this compound is a canonical example of a one-dimensional spin-1/2 AFHC. Interestingly, this weak interchain coupling in a bipartite lattice is due to the strong dimerization of the Cu2 ions. \cuox\ therefore represents a new route to extremely isolated AFHCs complementing structural isolation and interchain geometric frustration.

\begin{acknowledgments}
This work was supported by startup funding from the University of Central Florida. C.D., H.S., and Y.N. were supported by an NSF Career DMR-1944975. A.C.J. and B.J.P. were supported by the Australian Research Council, DP160100060. The work at Ames National Laboratory was supported by the U.S. Department of Energy, Division of Materials Sciences and Engineering. Ames National Laboratory is operated for the U.S. Department of Energy by Iowa State University under Contract No.~DE-AC02-07CH11358. The National High Magnetic Field Laboratory (NHMFL) is supported by the National Science Foundation through DMR-1644779 and the State of Florida. S.Y., X.H., and Y.T. were supported by the NHMFL UCGP Program.
\end{acknowledgments}

\bibliographystyle{apsrev4-2_YN.bst}
\end{document}